\documentclass[aps,pra,preprintnumbers,amsmath,amssymb,floatfix]{revtex4-2}

\usepackage{graphicx}
\usepackage{indentfirst}
\usepackage{color,soul}
\usepackage{braket}
\usepackage{tabularx}
\usepackage{hyperref}

\begin{document}

\title{Surviving Entanglement in Optic-Microwave Conversion\\by Electro-Optomechanical System}

\author{Yonggi Jo}\affiliation{Quantum Physics Technology Directorate, Advanced Defense Technology Research Institute, Agency for Defense Development, Yuseong P.O. Box 35, Daejeon 34186, Republic of Korea}
\author{Su-Yong Lee}\affiliation{Quantum Physics Technology Directorate, Advanced Defense Technology Research Institute, Agency for Defense Development, Yuseong P.O. Box 35, Daejeon 34186, Republic of Korea}
\author{Yong Sup Ihn}\affiliation{Quantum Physics Technology Directorate, Advanced Defense Technology Research Institute, Agency for Defense Development, Yuseong P.O. Box 35, Daejeon 34186, Republic of Korea}
\author{Dongkyu Kim}\affiliation{Quantum Physics Technology Directorate, Advanced Defense Technology Research Institute, Agency for Defense Development, Yuseong P.O. Box 35, Daejeon 34186, Republic of Korea}
\author{Zaeill Kim}\affiliation{Quantum Physics Technology Directorate, Advanced Defense Technology Research Institute, Agency for Defense Development, Yuseong P.O. Box 35, Daejeon 34186, Republic of Korea}
\author{Duk Y. Kim}\email[]{duk0@add.re.kr}\affiliation{Quantum Physics Technology Directorate, Advanced Defense Technology Research Institute, Agency for Defense Development, Yuseong P.O. Box 35, Daejeon 34186, Republic of Korea}

\date{\today}

\begin{abstract}
In recent development of quantum technologies, a frequency conversion of quantum signals has been studied widely. We investigate the optic-microwave entanglement that is generated by applying an electro-optomechanical frequency conversion scheme to one mode in an optical two-mode squeezed vacuum state. We quantify entanglement of the converted two-mode Gaussian state, where surviving entanglement of the state is analyzed with respect to the parameters of the electro-optomechanical system. Furthermore, we show that there exists an upper bound for the entanglement that survives after the conversion of highly entangled optical states. Our study provides a theoretical platform for a practical quantum illumination system.
\end{abstract}

\maketitle

\section{Introduction}

Entanglement is the essential source of a quantum advantage in quantum-enhanced sensing \cite{Giovannetti2011} and quantum illumination \cite{Lloyd2008,Tan2008,Barzanjeh2015}. In optics, we can generate two-mode entangled states with mean photon number of $7.4$ for each mode under the current technology \cite{Vahlbruch2016}. Although optical entanglement was demonstrated in a long-distance quantum communication over clear sky \cite{Ursin2007,Yin2020}, it is very challenging to distribute entanglement in free space due to its heavy scattering. To overcome the issue, it is natural to consider microwave regime which has lower attenuation than optical regime in the atmosphere. For example, $10$ GHz microwave signal that has very low attenuation, is used in radar system to detect a remote target. However, the microwave is in a low energy level so that thermal occupancy is a dominant issue.

To compensate both frequency regimes, we take both advantages of optical and microwave regimes, by performing optic-microwave bi-directional conversion. First, we prepare an optical entangled photon that is composed of signal and idler modes. Second, we convert the optical photon of the signal mode into microwave one. Third, we send the microwave photon into a target in the atmosphere while we keep the optical photon of the idler mode ideally. At this moment, we focus on how much entanglement the optic-microwave two-mode state contains. A frequency conversion of electromagnetic fields requires nonlinear interaction, which has been studied theoretically \cite{Vitali2007,Tsang2010,Tsang2011,OBrien2014,Everts2019}, and implemented experimentally in ferroelectric crystal \cite{Rueda2016,Rueda2019}, magnon \cite{Hisatomi2016, Ihn2020}, and an electro-optomechanical system \cite{Andrews2014,Higginbotham2018}. Up to now, the highest conversion efficiency was achieved with the electro-optomechanical system that consists of a microwave cavity(MC), an optical cavity(OC), and a mechanical resonator(MR). The MR that is coupled with the OC and the MC simultaneously, mediates a coherent conversion between them \cite{Barzanjeh2011}.

In this article, we analyze an optic-microwave entangled state generated by using the electro-optomechanical system. We start with an optical two-mode squeezed vacuum(TMSV) state which is conventionally exploited in continuous-variable quantum information processing. Then, the signal mode of the TMSV state is converted from an optical to a microwave frequency via the electro-optomechanical system while the idler mode is retained ideally. After the quantum frequency conversion, we analyze entanglement between the converted microwave signal mode and the optical idler mode. We find that entanglement can survive in the converted two-mode state when we employ a realistic system with feasible parameters \cite{Higginbotham2018}.

This article is organized as follows. In Sec.~\ref{SecSys} and \ref{SecConv}, we describe the electro-optomechanical system and its optic-microwave conversion process. In Sec.~\ref{SecQC}, we derive an entanglement formula between the converted signal mode and the idler mode, and analyze the amount of surviving entanglement compared with the entanglement of an input optical TMSV state. Then, we investigate a relation between the quantum frequency conversion and the amount of the surviving entanglement. Finally, it is concluded in Sec.~\ref{SecConclude}.

\section{Electro-optomechanical system}\label{SecSys}

\begin{figure}[ht!]
	\centering\includegraphics[width=0.7\textwidth]{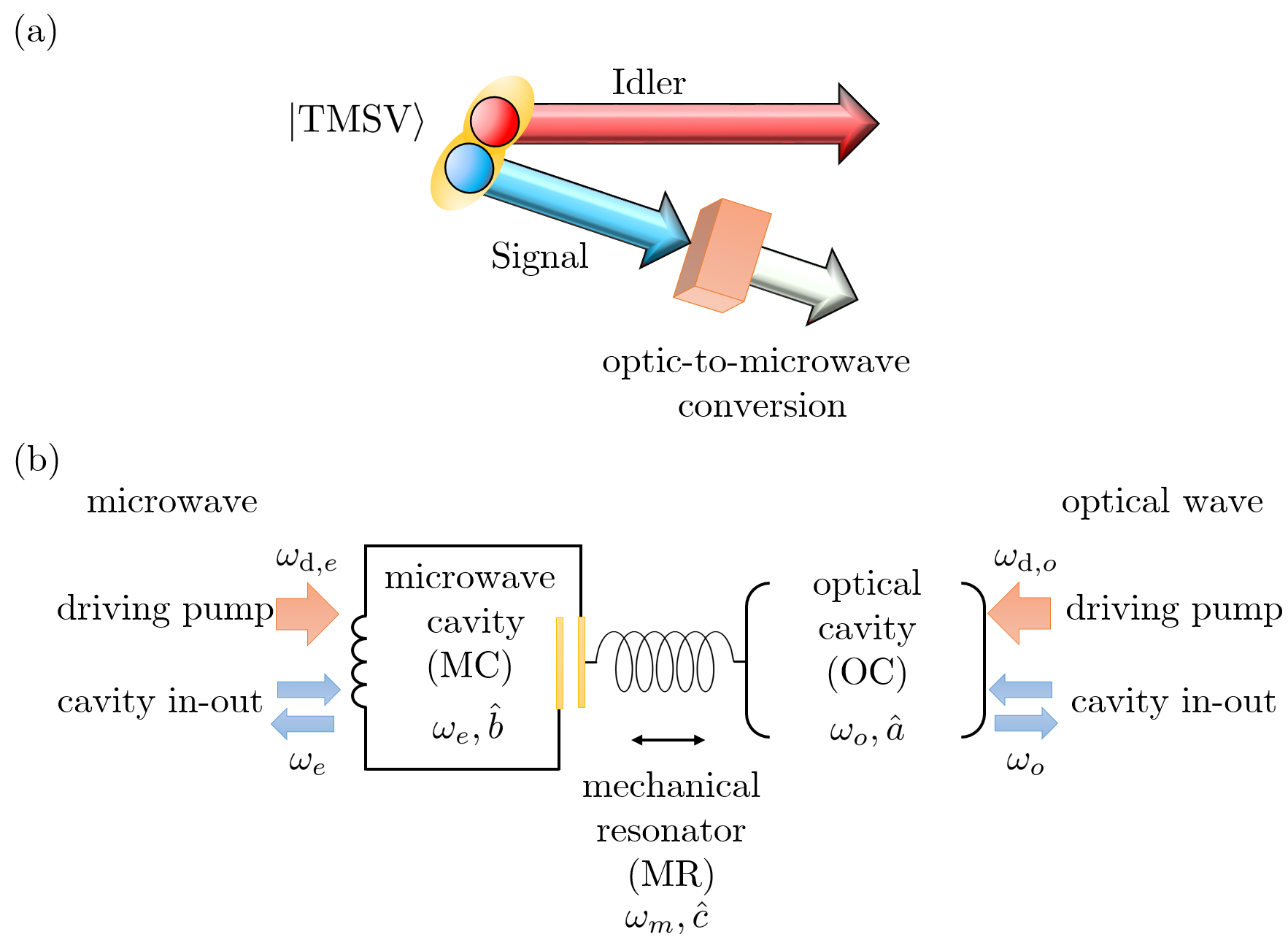}
	\caption{(a) A schematic diagram of an optic-microwave entangled state generation. One mode of a two-mode squeezed vacuum state is converted from an optical to microwave frequency while the other mode is ideally retained. (b) A schematic diagram of the optic-microwave conversion using electro-optomechanical system. A mechanical resonator connects an optical cavity with a microwave cavity, where $\omega_{m},~ \omega_{o},~ \omega_{e}$ are the resonance frequencies of MR, OC, MC, respectively.
The yellow lines denote a capacitor.
}\label{FigSys}
\end{figure}

Fig.~\ref{FigSys}(a) shows a schematic diagram of an optic-microwave entangled state generation. The signal mode of an optical TMSV state is converted from an optical to a microwave frequency, while the idler mode is ideally retained. Fig.~\ref{FigSys}(b) presents a schematic diagram of the electro-optomechanical system for the optic-microwave conversion. The MR is coupled to both the MC by the capacitance and the OC by modulating the optical pathway. The Hamiltonian of the entire system is written as
\begin{align}\label{totH}
\begin{split}
\hat{H}=&\hbar \omega_{o}\hat{a}^{\dagger}\hat{a}+\hbar\omega_{e}\hat{b}^{\dagger}\hat{b}+\hbar\omega_{m}\hat{c}^{\dagger}\hat{c}+\hbar g_{o}\hat{a}^{\dagger}\hat{a}(\hat{c}^{\dagger}+\hat{c})+\frac{\hbar g_{e}}{2}(\hat{b}^{\dagger}+\hat{b})^{2}(\hat{c}^{\dagger}+\hat{c})\\
&+i\hbar E_{o}(\hat{a}^{\dagger}e^{-i\omega_{\text{d},o}t}-\hat{a}e^{i\omega_{\text{d},o}t})+i\hbar E_{e}(e^{i\omega_{\text{d},e}t}-e^{-i\omega_{\text{d},e}t})(\hat{b}^{\dagger}+\hat{b}),
\end{split}
\end{align}
where $(\hat{a},\hat{a}^{\dagger})$, $(\hat{b},\hat{b}^{\dagger})$, and $(\hat{c},\hat{c}^{\dagger})$ show the annihilation and creation operators of the OC, MC, and MR, respectively $([\hat{a},\hat{a}^{\dagger}]=[\hat{b},\hat{b}^{\dagger}]=[\hat{c},\hat{c}^{\dagger}]=1)$ \cite{Barzanjeh2015}.
The driving pump frequency of the OC is set as $\omega_{\text{d},o}=\omega_{o}-\Delta_{\text{d},o}$, and the frequency of the MC driving pump is $\omega_{\text{d},e}=\omega_{e}-\Delta_{\text{d},e}$, where $\Delta_{\text{d},o}$ and $\Delta_{\text{d},e}$ denote optical and microwave pump detuning frequencies from the cavity resonance frequencies, respectively. 
 The subscripts $o$ and $e$ denote OC and MC. $g_{j}$ is the coupling strength, $E_{j}$ is the field strength of the input driving pump. The first three terms in Eq.~(\ref{totH}) denote the free Hamiltonian of the OC, MC, and MR. The fourth(or fifth) term shows an interaction between MR and OC(or MC). The last terms represent driving pumps of the OC and MC.

To analyze the optic-microwave conversion, we need to solve an input-output relation of the entire system, where the Hamiltonian of Eq.~(\ref{totH}) is simplified by linearizing the Hamiltonian \cite{Barzanjeh2011,Barzanjeh2015} and by the rotating wave approximation \cite{Allen1987,Barnett1997,Loudon2000}.
Then, the interaction Hamiltonian of Eq.~(\ref{totH}) can be linearized in the following equation:
\begin{align}
\begin{split}\label{intH}
\hat{H}_{I}=&\hbar G_{o}(\hat{c}^{\dagger}e^{i\omega_{m}t}+\hat{c}e^{-i\omega_{m}t})(\hat{\alpha}^{\dagger}e^{-i\Delta_{o}t}+\hat{\alpha}e^{-i\Delta_{o}t})\\
&+\hbar G_{e}(\hat{c}^{\dagger}e^{i\omega_{m}t}+\hat{c}e^{-i\omega_{m}t})(\hat{\beta}^{\dagger}e^{-i\Delta_{e}t}+\hat{\beta}e^{-i\Delta_{e}t}),
\end{split}
\end{align}
where $G_{j}=g_{j}\sqrt{N_{j}}$, $\hat{\alpha}\equiv\hat{a}-\sqrt{N_{o}}$, $\hat{\beta}\equiv\hat{b}-\sqrt{N_{e}}$ (intracavity quantum noise operators). $N_{j}=|E_{j}|^{2}/(\Gamma_{j}^{2}+\Delta_{j}^{2})$ is the mean photon number in the cavity, where $N_{j}\gg 1$ is induced by the driving pumps. The effective cavity detuning is obtained from the following equation:
\begin{align}
\Delta_{j}=\Delta_{\text{d},j}-g_{j}\frac{g_{o}N_{o}+g_{e}N_{e}}{\omega_{m}},
\end{align}
and $\Gamma_{j}=\gamma_{j}+\gamma_{j}'$ is a cavity decay rate which is the sum of an input loss rate $\gamma_{j}$ and an intrinsic loss rate $\gamma_{j}'$. 
The interaction Hamiltonian can be simplified with the appropriate setting of the effective cavity detuning. If both the optical and microwave pumps are red-detuned, i.e., $\omega_{m}=\Delta_{o}=\Delta_{e}$, then the interaction Hamiltonian can be simplified to
\begin{align}\label{reddetune}
\hat{H}_{I}=\hbar G_{o}(\hat{\alpha}\hat{c}^{\dagger}+\hat{\alpha}^{\dagger}\hat{c})+\hbar G_{e}(\hat{\beta}\hat{c}^{\dagger}+\hat{\beta}^{\dagger}\hat{c}),
\end{align}
under the rotating wave approximation, where the OC(or MC)-MR relation is a beam-splitter interaction. It means that one phonon in the MR is created when one optical(or microwave) photon is annihilated, and vice versa. Thus, one optical photon can be converted to one microwave photon through the MR.

\section{Optic-microwave conversion}\label{SecConv}

Under noise and damping, the dynamics of the electro-optomechanical system can be described by using quantum Langevin equation(QLE) \cite{Ford1988}. The nonlinear QLEs are written as follows:
\begin{subequations}\label{QLE}
\begin{align}
\dot{\hat{\alpha}}&=-\Gamma_{o}\hat{\alpha}-iG_{o}\hat{c}+\sqrt{2\gamma_{o}}\hat{\alpha}_{\text{in}}+\sqrt{2\gamma_{o}'}\hat{\alpha}_{\text{loss}},\\
\dot{\hat{\beta}}&=-\Gamma_{e}\hat{\beta}-iG_{e}\hat{c}+\sqrt{2\gamma_{e}}\hat{\beta}_{\text{in}}+\sqrt{2\gamma_{e}'}\hat{\beta}_{\text{loss}},\\
\dot{\hat{c}}&=-\gamma_{m}\hat{c}-iG_{o}\hat{\alpha}-iG_{e}\hat{\beta}+\sqrt{2\gamma_{m}}\hat{c}_{\text{loss}},
\end{align}
\end{subequations}
where $\gamma_{m}$ is a damping rate of the MR. The subscript ``in"(or ``out") denotes an input(or output) field, and ``loss" does intrinsic loss. 
Based on the cavity input-output relations \cite{Barzanjeh2011,Barzanjeh2015}:
\begin{align}\label{inout}
\begin{split}
\hat{\alpha}_{\text{out}}&=\sqrt{2\gamma_{o}}\hat{\alpha}-\hat{\alpha}_{\text{in}},\\
\hat{\beta}_{\text{out}}&=\sqrt{2\gamma_{e}}\hat{\beta}-\hat{\beta}_{\text{in}},
\end{split}
\end{align}
we derive an input-output relation between optical wave and microwave as:
\begin{align}\label{inoutrel}
\hat{\beta}_{\text{out}}=
C_{1}(\omega)\hat{\alpha}_{\text{in}}+ C_{2}(\omega)\hat{\beta}_{\text{in}}+ C_{3}(\omega)\hat{\alpha}_{\text{loss}}+ C_{4}(\omega)\hat{\beta}_{\text{loss}}+ C_{5}(\omega)\hat{c}_{\text{loss}},
\end{align}
where $\omega$ is the frequency difference of the output field from the resonance frequency. The coefficients of Eq.~(\ref{inoutrel}) are given as
\begin{subequations}\label{coeff}
	\begin{align}
		\label{inoutcoeff}
		C_{1}(\omega)&=-\frac{2G_{o}G_{e}\sqrt{\gamma_{o}\gamma_{e}}}{D(\omega)[G_{o}^{2}+(i\omega+\gamma_{m})(i\omega+\Gamma_{o})]},\\
		C_{2}(\omega)&=\frac{2\gamma_{e}}{D(\omega)}-1,\\
		C_{3}(\omega)&=-\frac{2G_{o}G_{e}\sqrt{\gamma_{o}'\gamma_{e}}}{D(\omega)[G_{o}^{2}+(i\omega+\gamma_{m})(i\omega+\Gamma_{o})]},\\
		C_{4}(\omega)&=\frac{2\sqrt{\gamma_{e}\gamma_{e}'}}{D(\omega)},\\
		C_{5}(\omega)&=-\frac{2i G_{e}\sqrt{\gamma_{e}\gamma_{m}}(i\omega+\Gamma_{o})}{D(\omega)[G_{o}^{2}+(i\omega+\gamma_{m})(i\omega+\Gamma_{o})]},
	\end{align}
\end{subequations}
where
\begin{align}\label{devider}
	D(\omega)=i\omega+\Gamma_{e}+\frac{G_{e}^{2}(i\omega+\Gamma_{o})}{G_{o}^{2}+(i\omega+\gamma_{m})(i\omega+\Gamma_{o})}.
\end{align}

Since we take the electro-optomechanical system in the recent demonstration \cite{Higginbotham2018}, we exploit the following values in our simulation: the coupling strengths are $g_{o}/2\pi=6.6$ Hz and $g_{e}/2\pi=3.8$ Hz; the OC and MC input loss rates are $2\gamma_{o}/2\pi=1.1$ MHz and $2\gamma_{e}/2\pi=2.3$ MHz; the intrinsic loss rates of OC and MC are $2\gamma_{o}'/2\pi=1$ MHz and $2\gamma_{e}'/2\pi=0.2$ MHz; the MR intrinsic loss rate $2\gamma_{m}/2\pi=11$ Hz; the resonance frequencies of OC, MC, and MR are $\omega_{o}/2\pi=282$ THz, $\omega_{e}/2\pi=6$ GHz, and $\omega_{m}/2\pi=1.4732$ MHz, respectively; the temperature of the electro-optomechanical system $T=35$ mK; and $N_{j}=1.7 \times 10^{8}$, where $j\in\{o,e\}$. Note that our loss rate is defined as half width at half maximum of the linewidth of the cavity \cite{Vitali2007,Barzanjeh2015}.

\section{Analysis of entanglement}\label{SecQC}
\subsection{Entanglement of converted two-mode Gaussian state}

A TMSV state is a conventional entangled state in continuous-variable quantum information processing. It can be generated by injecting two single-mode squeezed vacuum states into a 50:50 beam splitter \cite{Kim2002} or by using a spontaneous parametric down-conversion process \cite{Burnham1970,Takeoka2015}. The TMSV state can be written with the photon number basis as follows:
\begin{align}
\ket{\text{TMSV}}=\sum_{n=0}^{\infty}\sqrt{\frac{N_{S}^{n}}{(N_{S}+1)^{n+1}}}\ket{n}_{S}\ket{n}_{I},
\end{align}
where $N_{S}$ is the mean photon number of each mode, and the subscript $S$(or $I$) denotes the signal(or idler) mode of the TMSV state. As shown in Fig.~\ref{FigSys}(a), we analyze the quantum state obtained after converting from an optical to a microwave frequency in the signal mode of the TMSV state. 
Since thermal noise is produced by the electro-optomechanical system during the conversion, the output state is not a pure state but still in Gaussian regime.
We call it a converted two-mode Gaussian(CTMG) state 
which is given by the covariance matrix(CM) \cite{Vidal2002,Adesso2005} with $2\times 2$ block matrices A, B, and C as
\begin{align}
	V=\begin{bmatrix}
		A & C \\
		C^{\dagger} & B
	\end{bmatrix}.
\end{align}
The element of the CM can be obtained from
\begin{align}\label{covel}
	V_{ij}=\frac{1}{2}\braket{u_{i}u_{j}+u_{j}u_{i}},
\end{align}
where 
\begin{align}
	\vec{u}=\left[\hat{x}_{S},\hat{p}_{S},\hat{x}_{I},\hat{p}_{I}\right]^{T},
\end{align}
and $\hat{x}_{S}=(\hat{a}_{S}^{\dagger}+\hat{a}_{S})/\sqrt{2}$, $\hat{p}_{S}=i(\hat{a}_{S}^{\dagger}-\hat{a}_{S})/\sqrt{2}$, $\hat{x}_{I}=(\hat{a}_{I}^{\dagger}+\hat{a}_{I})/\sqrt{2}$, $\hat{p}_{I}=i(\hat{a}_{I}^{\dagger}-\hat{a}_{I})/\sqrt{2}$.

Entanglement of the CTMG can be quantified by the logarithmic negativity(LN) \cite{Audenaert2003} that is evaluated from the symplectic eigenvalues under partial transposition. If a symplectic eigenvalue of a partially transposed two-mode Gaussian state are smaller than $1/2$, the two-mode Gaussian state is entangled \cite{Weedbrook2012}.
The symplectic eigenvalues under partial transposition can be obtained from the following equation \cite{Tahira2009}:
\begin{align}
	\xi^{4}-(\text{Det}[A]+\text{Det}[B]-2\text{Det}[C])\xi^{2}+\text{Det}[V]=0,
\end{align}
where Det$[X]$ denotes the determinant of matrix $X$. From two positive roots of this equation $\xi_{\pm}$, the LN is defined as
\begin{align}
	\text{LN}\equiv\max\{0,-\text{ln}[2\xi_{-}]\},
\end{align}
where $\xi_{+}$ always satisfies $\xi_{+}\geq 1/2$ \cite{Tahira2009}.
Since $\text{Det}[A]=\text{Det}[B]$ for any TMSV state, the LN of an input TMSV state is given by
\begin{align}\label{TMSVLN}
	\text{LN}_{\text{TMSV}}=-\ln \left[2N_{S}+1-2\sqrt{N_{S}(N_{S}+1)}\right].
\end{align}

Now we calculate the LN of the CTMG state.  Since we consider loss and noise in our QLE, we take a thermal bath as loss and noise in the conversion.
Initially we prepare a TMSV state with the thermal bath as
\begin{align}
	\hat{\rho}_{\text{total}}=\ket{\text{TMSV}}_{SI}\bra{\text{TMSV}}\otimes \hat{\rho}_{\text{Th}},
\end{align}
where $\hat{\rho}_{\text{Th}}$ is the thermal bath.
Using the microwave-converted signal mode operator that is the microwave output mode $\hat{\beta}_{S}=\hat{\beta}_{\text{out}}$ of Eq.~(\ref{inoutrel}), we can derive the expectation values under the input-output relation in Eq.~(\ref{inoutrel}). For example,
\begin{subequations}\label{corr}
	\begin{align}
		\begin{split}\label{corr1}
			\braket{\hat{\beta}^{\dagger}_{S}\hat{\beta}_{S}}=&|C_{1}(\omega)|^{2}\braket{\hat{\alpha}_{\text{in}}^{\dagger}\hat{\alpha}_{\text{in}}}+|C_{2}(\omega)|^{2}\braket{\hat{\beta}_{\text{in}}^{\dagger}\hat{\beta}_{\text{in}}}+|C_{3}(\omega)|^{2}\braket{\hat{\alpha}_{\text{loss}}^{\dagger}\hat{\alpha}_{\text{loss}}}\\
			&+|C_{4}(\omega)|^{2}\braket{\hat{\beta}_{\text{loss}}^{\dagger}\hat{\beta}_{\text{loss}}}+|C_{5}(\omega)|^{2}\braket{\hat{c}_{\text{loss}}^{\dagger}\hat{c}_{\text{loss}}},
		\end{split}\\
		\braket{\hat{a}_{I}^{\dagger}\hat{a}_{I}}=&N_{S},\\
		\braket{\hat{\beta}_{S}\hat{a}_{I}}=&C_{1}(\omega)\braket{\hat{\alpha}_{\text{in}}\hat{a}_{I}}.\label{psc}
	\end{align}
\end{subequations}
The block matrices $A$, $B$, and $C$ can be obtained from the expectation values. 
In the optical input mode of the electro-optomechanical system, if we ignore the stationary cavity field, there are the signal mode of the TMSV state and a thermal noise, resulting in $\braket{\hat{\alpha}_{\text{in}}^{\dagger}\hat{\alpha}_{\text{in}}}=N_{S}+\widetilde{N}_{o}$. 
 The microwave input mode and the loss modes are not related with the input TMSV state, such that the other expectation values are given by  $\braket{\hat{\beta}_{\text{in}}^{\dagger}\hat{\beta}_{\text{in}}}=\widetilde{N}_{e}$,  $\braket{\hat{\alpha}_{\text{loss}}^{\dagger}\hat{\alpha}_{\text{loss}}}=\widetilde{N}_{o}$, $\braket{\hat{\beta}_{\text{loss}}^{\dagger}\hat{\beta}_{\text{loss}}}=\widetilde{N}_{e}$, and $\braket{\hat{c}_{\text{loss}}^{\dagger}\hat{c}_{\text{loss}}}=\widetilde{N}_{m}$. 
Note that $\widetilde{N}_{j}=[\exp(\hbar \omega_{j}/k_{B}T)-1]^{-1}$ is the mean photon number of the thermal noise induced from the electro-optomechanical system and $T$ is a temperature of the electro-optomechanical system. In the phase-sensitive cross correlation of Eq.~(\ref{psc}), only $\hat{\alpha}_{\text{in}}$ mode has a correlation with the idler mode whereas the other terms become zero. The input optical wave is the signal mode of the TMSV state with thermal noise which is not related with the idler mode, resulting in $\braket{\hat{\alpha}_{\text{in}}\hat{a}_{I}}=\sqrt{N_{S}(N_{S}+1)}$.

A full description of $\xi_{-}$ of the CTMG state can be derived analytically. We show $\xi_{-}$ at $\omega=0$ as follows:
\begin{align}\label{symeigen}
	\xi_{-}|_{\omega=0}=\frac{1}{4}\left[d_{1}+d_{2}-\sqrt{(d_{1}-d_{2})^{2}+4d_{3}^{2}}\right],
\end{align}
where
\begin{subequations}\label{lncoefsimple}
\begin{align}
		d_{1}=& \frac{8G_{o}^{2}G_{e}^{2}\gamma_{o}\gamma_{e}}{Z^{2}}N_{S}+1+2\widetilde{N}_{e}+\frac{8G_{e}^{2}\Gamma_{o}\gamma_{e}}{Z^{2}}\left[G_{o}^{2}\widetilde{N}_{o}-(G_{o}^{2}+\Gamma_{o}\gamma_{m})\widetilde{N}_{e}+\Gamma_{o}\gamma_{m}\widetilde{N}_{m}]\right],\\
		d_{2}=&2N_{S}+1,\\
		d_{3}=&\frac{4G_{o}G_{e}\sqrt{\gamma_{o}\gamma_{e}}}{Z}\sqrt{N_{S}(N_{S}+1)},
\end{align}
\end{subequations}
with
\begin{align}
	Z=G_{o}^{2}\Gamma_{e}+G_{e}^{2}\Gamma_{o}+\Gamma_{o}\Gamma_{e}\gamma_{m}.
\end{align}
We calculate numerically the amount of surviving entanglement of the CTMG state which is compared with entanglement of the initial TMSV state, Eq.~(\ref{TMSVLN}). 

\begin{figure}[t!]
	\centering\includegraphics[width=0.95\textwidth]{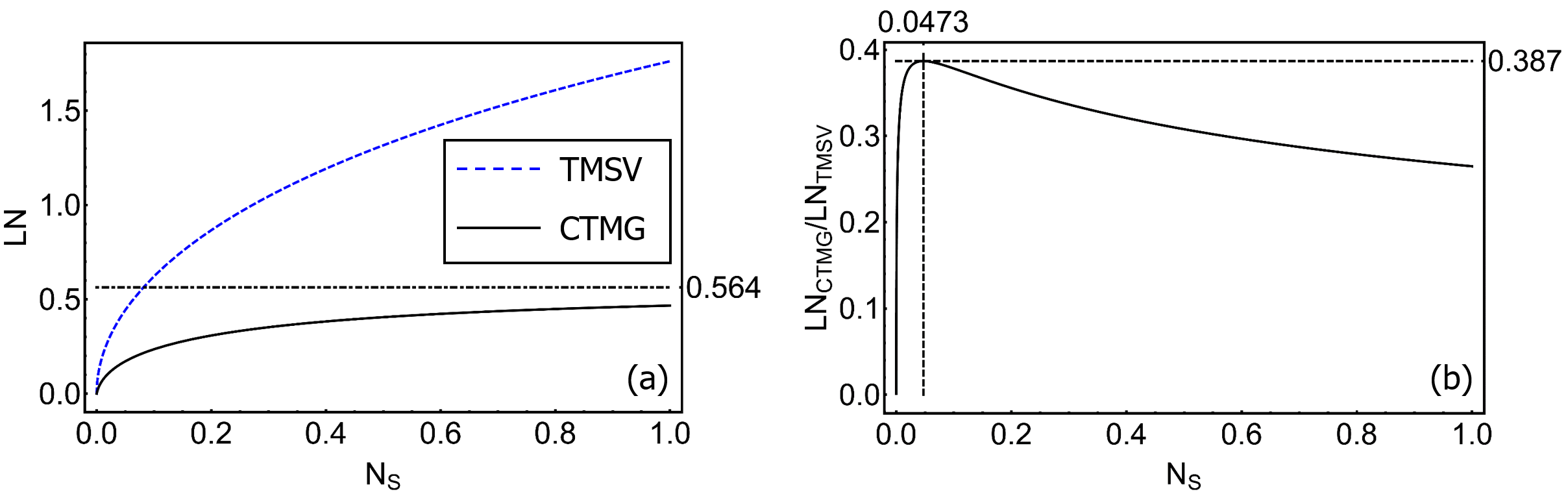}
	\caption{(a) LNs of the CTMG state(black solid line) and the input TMSV state(blue dashed line) as a function of the mean photon number of the signal mode in the initial TMSV state. The black dot-dashed line at $\text{LN}=0.564$ denotes the asymptotic limit of the LN of the CTMG state. (b) Entanglement surviving ratio after the optic-microwave conversion. The maximum entanglement surviving ratio is approximately $0.387$ at $N_{S}\approx 0.047$ (black dashed lines).}\label{FigTMSV}
\end{figure}

Fig.~\ref{FigTMSV}(a) shows the LNs of the input TMSV state and the CTMG state as a function of the mean photon number($N_{S}$) of the signal mode in the TMSV state. The entanglement of the input TMSV state partially survives after the optic-microwave conversion of the signal mode. The LN of the CTMG state has non-zero values and increases up to $0.564$. In the first proposal of Gaussian quantum illumination \cite{Tan2008}, there is quantum advantage when a TMSV state with $N_{S}=0.01$ is exploited, where the LN of the state is roughly 0.2. Thus, we expect that our CTMG state can be exploited in quantum illumination.

Fig.~\ref{FigTMSV}(b) shows the entanglement surviving ratio after the optic-microwave conversion as a function of $N_{S}$. Under a fixed conversion efficiency, the entanglement survival of the CTMG state depends on the system of the frequency conversion as well as the injected input mean photon number. With our electro-optomechanical system, the entanglement surviving ratio between the converted state and  the input TMSV state is maximized as approximately $0.387$ at $N_{S}=0.047$. Due to the induced thermal noise, the entanglement of the initial state does not well survive in the very low limit of $N_{S}$.

The LN of the TMSV state in Eq.~(\ref{TMSVLN}) can be infinitely large with increasing $N_{S}$, but the LN of the CTMG state is upper bounded by $0.564$. Since the upper bound depends on the system parameters, we define this upper bound as the entanglement surviving capacity of the electro-optomechanical system.  The capacity is defined as the LN of the CTMG state in the limit of $N_{S}\rightarrow \infty$:
\begin{align}
	\text{LN}_{\text{CTMG}}\leq P\equiv&\max\{0,-\lim_{N_{S}\rightarrow\infty}\text{ln}\left[k_{1}+k_{2}N_{S}-\sqrt{k_{3}+k_{4}N_{S}+k_{5}N_{S}^{2}}\right]\},
\end{align}
where $k_{i}$ with $i\in\{1,2,3,4,5\}$ is given in appendix. The entanglement surviving capacity $P$ converges if $k_{2}=(k_{5})^{2}$ satisfies, which is always true in our system. By taking the Taylor expansion, the capacity becomes
\begin{align}\label{capa}
	P\sim \max\{0,-\text{ln}\left[k_{1}-\frac{k_{4}}{2k_{2}}\right]\}.
\end{align}
If we consider the ideal situation, i.e., the noiseless conversion at $T\rightarrow 0$, the capacity is simplified as:
\begin{align}\label{PTzero}
	\begin{split}
	P|_{T\rightarrow 0}\sim &\max\{0,-\text{ln}\left[1-\frac{8G_{o}^{2}G_{e}^{2}\gamma_{o}\gamma_{e}}{Z^{2}+4G_{o}^{2}G_{e}^{2}\gamma_{o}\gamma_{e}}\right]\}\\
	=&\max\{0,-\text{ln}\left[1-\frac{2R(0)}{1+R(0)}\right]\},
	\end{split}
\end{align}
where $R(0)$ is the frequency conversion efficiency of the electro-optomechanical system at $\omega=0$, which is written as
\begin{align}\label{maxeff}
	R(0)=\frac{4G_{o}^{2}G_{e}^{2}\gamma_{o}\gamma_{e}}{Z^{2}}.
\end{align}
From Eq.~(\ref{PTzero}), we find that the entanglement surviving capacity under ideal condition is directly related with the conversion efficiency of the system. It is more interpreted in the next section.

\subsection{Conversion efficiency and entanglement survival}

We analyze the conversion efficiency of the electro-optomechanical system whose optimal condition is compared with the optimal one on the surviving entanglement of the CTMG state. From Eq.~(\ref{inoutrel}), we define an optic-microwave conversion efficiency as $R(\omega)\equiv |C_{1}(\omega)|^{2}$, the probability that an input optical photon is converted to a microwave photon \cite{Tsang2011}. 
Since $C_{1}(\omega)$ of Eq.~(\ref{inoutcoeff}) does not depend on the number of input photons or temperature of the conversion system, 
$R(\omega)$ represents a characteristic of the electro-optomechanical system. 
At $\omega=0$, the conversion efficiency of our system approaches to $R(0)\approx 0.328$, so that less than a half of input optical photons are converted to microwave photons.
It is compared with the existing experiments \cite{Andrews2014,Higginbotham2018} whose conversion efficiency is defined as the ratio between input amplitude and converted amplitude, $\braket{\hat{\beta}_{S}^{\dagger}\hat{\beta}_{S}}/\braket{\hat{a}_{S}^{\dagger}\hat{a}_{S}}$ in our calculation. Due to the added thermal noise, their conversion efficiency varies with the input mean photon number and the temperature of the electro-optomechanical system.  Thus, if we apply our system to their calculation, our conversion efficiency can be larger than $R(0)\approx 0.328$, such as $\braket{\hat{\beta}_{S}^{\dagger}\hat{\beta}_{S}}/\braket{\hat{a}_{S}^{\dagger}\hat{a}_{S}}=0.482,~0.636$ at $N_{S}=1,~0.5$, respectively.

\begin{figure}[t!]
	\centering\includegraphics[width=0.9\textwidth]{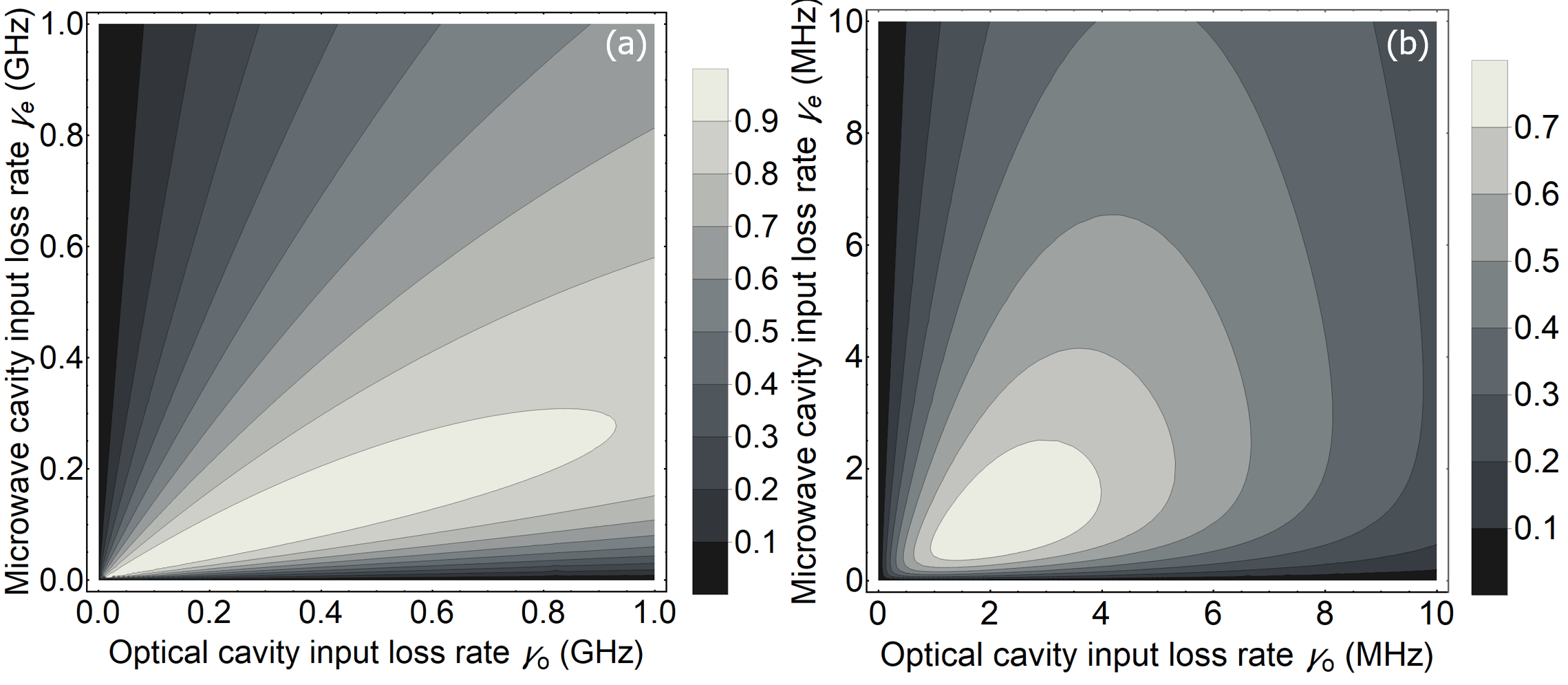}
	\caption{At $N_{S}=1$, $T=35$ mK, and $\omega=0$, (a) optic-microwave conversion efficiency of electro-optomechanical system and (b) LN of the CTMG state as a function of the two cavity loss rates. }\label{FigCE_LN}
\end{figure}

If we can control the cavity input loss rates with other parameters fixed, the conversion efficiency is maximized at the following values:
\begin{subequations}
	\begin{align}\label{maxinloss}
		\gamma_{o}|_{\text{max}[R]}&=\sqrt{\frac{(G_{o}^{2}+\gamma_{o}'\gamma_{m})\left[G_{e}^{2}\gamma_{o}'+(G_{o}^{2}+\gamma_{o}'\gamma_{m})\gamma_{e}'\right]}{(G_{e}^{2}+\gamma_{e}'\gamma_{m})\gamma_{m}}},\\
		\gamma_{e}|_{\text{max}[R]}&=\sqrt{\frac{(G_{e}^{2}+\gamma_{e}'\gamma_{m})\left[G_{e}^{2}\gamma_{o}'+(G_{o}^{2}+\gamma_{o}'\gamma_{m})\gamma_{e}'\right]}{(G_{o}^{2}+\gamma_{o}'\gamma_{m})\gamma_{m}}},
	\end{align}
\end{subequations}
where the cavity input loss rates are related with the coupling rates and the intrinsic loss rates. In our electro-optomechanical system, the conversion efficiency is maximized as $0.962$ at $\gamma_{o}=82.3$ MHz and $\gamma_{e}=27.3$ MHz, as shown in Fig.~\ref{FigCE_LN}(a).
It shows $R(0)$ as a function of the optical and microwave cavity input loss rates. The other parameters are the same as those described in the Sec.~\ref{SecConv}. As previously studied \cite{Andrews2014,Higginbotham2018}, the conversion efficiency maximizes when $G_{o}^{2}/\Gamma_{o}\gamma_{m}$ and $G_{e}^{2}/\Gamma_{e}\gamma_{m}$ are equivalent, whereas it gets smaller when the values are far apart. 

Fig.~\ref{FigCE_LN}(b) shows the LN of the CTMG state with respect to the cavity input loss rates at $\omega=0$ when the input TMSV state satisfies $N_{S}=1$. Note that the units of the cavity input loss rates are GHz in Fig.~\ref{FigCE_LN}(a) and MHz in Fig.~\ref{FigCE_LN}(b). 
The maximum LN of the CTMG state is obtained at $\gamma_{o}=1.82$ MHz and $\gamma_{e}=0.818$ MHz while the conversion efficiency with these input loss rates is $0.694$. It is different from the optimal condition of the conversion effficiency which is obtained at the cavity input loss rates $\gamma_{o}=82.3$ MHz and $\gamma_{e}=27.3$ MHz.
These results implicate that the higher conversion efficiency does not guarantee the better survival of entanglement. Although we showed that the maximum entanglement surviving capacity is related with the conversion efficiency of the system as shown in Eq.~(\ref{PTzero}), their optimal conditions are differed by the thermal noise induced from the conversion system. The thermal noise is inevitable in realistic system at finite temperature. Therefore, we can conclude that in order to enhance the entanglement survival, the system should be designed not for maximizing the conversion efficiency, but for maximizing the entanglement surviving capacity with the consideration of the system temperature.

\section{Conclusion and discussion}\label{SecConclude}

In this article, we analyzed entanglement of an optic-microwave entangled state that is generated by applying the quantum frequency conversion with an electro-optomechanical system to one mode in an optical two-mode squeezed vacuum(TMSV) state. We investigated entanglement of the converted two-mode Gaussian(CTMG) state by the logarithmic negativity(LN), which is compared with the input TMSV state. 
We showed that the entanglement of the CTMG state has an upper bound which is a characteristic of the frequency conversion system.
It was also found that the entanglement surviving ratio is affected by the mean photon number of the input TMSV state, even if the conversion efficiency of the electro-optomechanical system is fixed. 
Finally, we obtained that the higher conversion efficiency does not guarantee the larger entanglement of the CTMG state, even though the conversion efficiency is directly related with the entanglement surviving capacity at the zero temperature limit. 
Thus, in order to generate a highly entangled optic-microwave two-mode state, we have to adjust the conversion system parameters for enhancing the entanglement survival rather than for the conversion efficiency.

Under the same electro-optomechanical system, an optic-microwave entangled state is directly generated by using one red-detuned pump and one blue-detuned pump in \cite{Barzanjeh2015}, whereas our scheme exploits two red-detuned pumps to convert a signal mode from optical regime to microwave one. Compared to the direct generation scheme, we may infer intuitively that our scheme has advantages over controlling properties of the output state. For example, as shown in Fig.~\ref{FigTMSV}(a), the entanglement of the CTMG state can be controlled simply by changing the mean photon number of the optical TMSV state rather than modifying the whole electro-optomechanical system. We leave the detail comparison as a further work.

From the results, we theoretically confirm that the input state entanglement can survive after the quantum frequency conversion, and the CTMG state can be used in a long-distance target detection and quantum network that exploit optic-microwave entanglement.

\section*{Appendix}
\begin{appendix}
	\section{Coefficients of entanglement surviving capacity}
	
	Here, we describe the coefficients $k_{i}$ with $i\in\{1,2,3,4,5\}\}$ for the entanglement surviving capacity in Eq.~(\ref{capa}). The values are following:
	\begin{align}\nonumber
		\begin{split}
			k_{1}=&1+\widetilde{N}_{e}+\frac{4G_{e}^{2}\Gamma_{o}\gamma_{e}}{Z^{2}}\left[G_{o}^{2}\widetilde{N}_{o}-(G_{o}^{2}+\Gamma_{o}\gamma_{m})\widetilde{N}_{e}+\gamma_{m}\widetilde{N}_{m}\right],\\
			k_{2}=&1+\frac{4G_{o}^{2}G_{e}^{2}\gamma_{o}\gamma_{e}}{Z^{2}},\\
			k_{3}=&\left\{\widetilde{N}_{e}+\frac{4G_{e}^{2}\Gamma_{o}\gamma_{e}}{Z^{2}}\left[G_{o}^{2}\widetilde{N}_{o}-(G_{o}^{2}+\Gamma_{o}\gamma_{m})\widetilde{N}_{e}+\gamma_{m}\widetilde{N}_{m}\right]\right\}^{2},\\
			k_{4}=&\frac{2}{Z^{4}}\left(l_{1}\widetilde{N}_{o}+l_{2}\widetilde{N}_{e}+l_{3}\widetilde{N}_{m}+l_{4}\right),\\
			k_{5}=&1+\frac{8G_{o}^{2}G_{e}^{2}\gamma_{o}\gamma_{e}}{Z^{2}}+\frac{16G_{o}^{4}G_{e}^{4}\gamma_{o}^{2}\gamma_{e}^{2}}{Z^{4}},
		\end{split}
	\end{align}
	where
	\begin{align}\nonumber
		\begin{split}
			l_{1}&=-4G_{o}^{2}G_{e}^{2}\Gamma_{o}\gamma_{e}(Z^{2}-4G_{o}^{2}G_{e}^{2}\gamma_{o}\gamma_{e}),\\
			l_{2}&=-Z^{4}+4G_{e}^{2}\gamma_{e}\left[G_{o}^{2}(\Gamma_{o}+\gamma_{o})+\Gamma_{o}^{2}\gamma_{m})\right]Z^{2}-16G_{o}^{2}G_{e}^{4}\Gamma_{o}\gamma_{o}\gamma_{e}^{2}(G_{o}^{2}+\Gamma_{o}\gamma_{m}),\\
			l_{3}&=-4G_{e}^{2}\Gamma_{o}^{2}\gamma_{e}\gamma_{m}(Z^{2}-4G_{o}^{2}G_{e}^{2}\gamma_{o}\gamma_{e}),\\
			l_{4}&=8G_{o}^{2}G_{e}^{2}\gamma_{o}\gamma_{e}Z^{2},
		\end{split}
	\end{align}
	with $Z=G_{o}^{2}\Gamma_{e}+G_{e}^{2}\Gamma_{o}+\Gamma_{o}\Gamma_{e}\gamma_{m}$. If $N_{S}=0$, the entanglement surviving capacity becomes $P=-\text{ln}\left[k_{1}-\sqrt{k_{3}}\right]=0$ since $k_{1}-\sqrt{k_{3}}=1$ satisfies. The noiseless entanglement surviving capacity $P|_{T\rightarrow 0}$ converges since $k_{2}=(k_{5})^{2}$ satisfies.
	
\end{appendix}

\section*{Acknowledgments}
We would like to thank Donghwan Kim for constructive comments on the manuscript. This work was supported by a grant to the Quantum Standoff Sensing Defense-Specialized Project funded by the Defense Acquisition Program Administration and the Agency for Defense Development.


\begin{thebibliography}{1}
\newcommand{\enquote}[1]{``#1''}

\bibitem{Giovannetti2011} V. Giovannetti, S. Lloyd, and L. Maccone, \enquote{Advances in quantum metrology,} Nat. Photon. \textbf{5}, pp. 222--229 (2011).

\bibitem{Lloyd2008} S. Lloyd, \enquote{Enhanced Sensitivity of Photodetection via Quantum Illumination,} Science {\bf 321}, pp. 1463--1465 (2008).

\bibitem{Tan2008} S.-H. Tan, B. I. Erkmen, V. Giovannetti, S. Guha, S. Lloyd, L. Maccone, S. Pirandola, and J. H. Shapiro, \enquote{Quantum Illumination with Gaussian States,} Phys. Rev. Lett. {\bf 101}, 253601 (2008).

\bibitem{Barzanjeh2015} Sh. Barzanjeh, S. Guha, C. Weedbrook, D. Vitali, J. Shapiro, and S. Pirandola, \enquote{Microwave Quantum Illumination,} Phys. Rev. Lett. {\bf 114}, 080503 (2015).

\bibitem{Vahlbruch2016} H. Vahlbruch, M. Mehmet, K. Danzmann, and R. Schnabel, \enquote{Detection of 15 dB Squeezed States of Light and their Application for the Absolute Calibration of Photoelectric Quantum Efficiency,} Phys. Rev. Lett. \textbf{117}, 110801 (2016).

\bibitem{Ursin2007} R. Ursin, F. Tiefenbacher, T. Schmitt-Manderbach, H. Weier, T. Scheidl, M. Lindenthal, B. Blauensteiner, T. Jennewein, J. Perdigues, P. Trojek, B. \"{O}mer, M. F\"{u}rst, M. Meyenburg, J. Rarity, Z. Sodnik, C. Barbieri, H. Weinfurter, and A. Zeilinger, \enquote{Entanglement-based quantum communication over 144km,} Nat. Phys. {\bf 3}, pp. 481--486 (2007).

\bibitem{Yin2020} J. Yin, Y.-H. Li, S.-K. Liao, M. Yang, Y. Cao, L. Zhang, J.-G. Ren, W.-Q. C, W.-Y. Liu, S.-L. Li, R. Shu, Y.-M. Huang, L. Deng, L. Li, Q. Zhang, N.-L. Liu, Y.-A. Chen, C.-Y. Lu, X.-B. Wang, F. Xu, J.-Y. Wang, C.-Z. Peng, A. K. Ekert, and J.-W. Pan, \enquote{Entanglement-based secure quantum cryptography over 1,120 kilometres,} Nature {\bf 582}, pp. 501--505 (2020).

\bibitem{Vitali2007} D. Vitali, P. Tombesi, J. J. Woolley, A. C. Doherty, and G. J. Milburn, \enquote{Entangling a nanomechanical resonator and a superconducting microwave cavity,} Phys. Rev. A {\bf 76}, 042336 (2007).

\bibitem{Tsang2010} M. Tsang, \enquote{Cavity quantum electro-optics,} Phys. Rev. A {\bf 81}, 063837 (2010).

\bibitem{Tsang2011} M. Tsang, \enquote{Cavity quantum electro-optics. II. Input-output relations between traveling optical and microwave fields,} Phys. Rev. A {\bf 84}, 043845 (2011).

\bibitem{OBrien2014} C. O'Brien, N. Lauk, S. Blum, G. Morigi, and M. Fleischhauer, \enquote{Interfacing Superconducting Qubits an Telecom Photons via a Rare-Earth-Doped Crystal,} Phys. Rev. Lett. {\bf 113}, 063603 (2014).

\bibitem{Everts2019} J. R. Everts, M. C. Berrington, R. L. Ahlefeldt, and J. J. Longdell, \enquote{Microwave to optical photon copnversion via fully concentrated rare-earth-ion crystals,} Phys. Rev. A {\bf 99}, 063830 (2019).

\bibitem{Rueda2016} A. Rueda, F. Sedlmeir, M. C. Collodo, U. Vogl, B. Stiller, G. Schunk, D. V. Strekalov, C. Marquardt, J. M. Fink, O. Painter, G. Leuchs, and H. G. L. Schwefel, \enquote{Efficient microwave to optical photon conversion: an electro-optical realization,} Optica {\bf 3}(6), pp. 597--604 (2016).

\bibitem{Rueda2019} A. Rueda, W. Hease, Sh. Barzanjeh, and J. M. Fink, \enquote{Electro-optic entanglement source for microwave to telecom quantum state transfer,} npj Quantum Inf. {\bf 5}, 108 (2019).

\bibitem{Hisatomi2016} R. Hisatomi, A. Osada, Y. Tabuchi, T. Ishikawa, A. Noguchi, R. Yamazaki, K. Usami, and Y. Nakamura, \enquote{Bidirectional conversion between microwave an light via ferromagnetic magnons,} Phys. Rev. B {\bf 93}, 174427 (2016).

\bibitem{Ihn2020} Y. S. Ihn, S.-Y. Lee, D. Kim, S. H. Yim, and Z. Kim, \enquote{Coherent multimode conversion from microwave to optical wave via a magnon-cavity hybrid system,} Phys. Rev. B {\bf 102}, 064418 (2020).


\bibitem{Andrews2014} R. W. Andrews, R. W. Peterson, T. P. Purdy, K. Cicak, R. W. Simmonds, C. A. Regal, and K. W. Lehnert, \enquote{Bidirectional and efficient conversion between microwave and optical light,} Nat. Phys. {\bf 10}, pp. 321--326 (2014).

\bibitem{Higginbotham2018} A. P. Higginbotham, P. S. Burns, M. D. Urmey, R. W. Peterson, N. S. Kampel, B. M. Brubaker, G. Smith, K. W. Lehnert, and C. A. Regal, \enquote{Harnessing electro-optic correlations in an efficient mechanical converter,} Nat. Phys. \textbf{14}, pp. 1038--1042 (2018).

\bibitem{Barzanjeh2011} Sh. Barzanjeh, D. Vitali, P. Tombesi, and G. J. Milburn, \enquote{Entangling optical and microwave cavity modes by means of a nanomechanical resonator,} Phys. Rev. A {\bf 84}, 042342 (2011).

\bibitem{Allen1987} L. Allen and J. H. Eberly, \enquote{Optical Resonance and Two-Level Atoms} (Dover Publications, New York, 1987).
	
\bibitem{Barnett1997} S. M. Barnett and P. M. Radmore, \enquote{Methods in Theortical Quantum Optics} (Oxford Science Publications, New York, 1997).
		
\bibitem{Loudon2000} R. Loudon, \enquote{The Quantum Theory of Light} (Oxford Science Publications, New York, 2000).
			
\bibitem{Ford1988} G. W. Ford, J. T. Lewis, and R. F. O'Connell, \enquote{Quantum Langevin equation,} Phys. Rev. A {\bf 37}(11), 4419 (1988).

\bibitem{Kim2002} M. S. Kim, W. Son, V. Bu{\u z}ek, and P. L. Knight, \enquote{Entanglement by a beam splitter: Nonclassicality as a prerequisite for entanglement,} Phys. Rev. A {\bf 65}, 032323 (2002).

\bibitem{Burnham1970} D. C. Burnham and D. L. Weinberg, \enquote{Observation of Simultaneity in Parametric Production of Optical Photon Pairs,} Phys. Rev. Lett. {\bf 25}, 84 (1970).

\bibitem{Takeoka2015} M. Takeoka, R.-B. Jin, and M. Sasaki, \enquote{Full analysis of multi-photon pair effects in spontaneous parametric down conversion based photonic quantum information processing,} New J. Phys. {\bf 17}, 043030 (2015).





\bibitem{Vidal2002} G. Vidal and R. F. Werner, \enquote{Computable measure of entanglement,} Phys. Rev. A {\bf 65}, 032314 (2002).

\bibitem{Adesso2005} G. Adesso and F. Illuminati, \enquote{Gaussian measures of entanglement versus negativities: Ordering of two-mode Gaussian states,} Phys. Rev. A {\bf 72}, 032334 (2005).


\bibitem{Audenaert2003} K. Audenaert, M. B. Plenio, and J. Eisert, \enquote{Entanglement Cost under Positive-Partial-Transpose-Preserving Operations,} Phys. Rev. Lett. {\bf 90}, 027901 (2003).

\bibitem{Weedbrook2012} C. Weedbrook, S. Pirandola, R. Gar{\' i}a-Patr{\' o}n, N. J. Cerf, T. C. Ralph, J. H. Shapiro, and S. Lloyd, \enquote{Gaussian quantum information,} Rev. Mod. Phys. {\bf 84}, pp. 621--669 (2012).

\bibitem{Tahira2009} R. Tahira, M. Ikram, H. Nha, and M. S. Zubairy, \enquote{Entanglement of Gaussian states using a beam splitter,} Phys. Rev. A {\bf 79}, 023816 (2009).


\end{thebibliography}
\end{document}